\documentclass[aip, reprint, amsmath, amssymb, floatfix]{revtex4-1}

\usepackage[utf8]{inputenc}
\usepackage[T1]{fontenc}
\usepackage{mathptmx}
\usepackage{etoolbox}
\usepackage{amsmath}
\usepackage{siunitx}
\usepackage{comment}
\usepackage{graphicx}
\usepackage{dcolumn}% Align table columns on decimal point
\usepackage{bm}% bold math
\usepackage{amssymb, amsmath, amsbsy,amsthm}
\usepackage{multibib}
\makeatletter
\def\@email#1#2{%
 \endgroup
 \patchcmd{\titleblock@produce}
  {\frontmatter@RRAPformat}
  {\frontmatter@RRAPformat{\produce@RRAP{*#1\href{mailto:#2}{#2}}}\frontmatter@RRAPformat}
  {}{}
}%
\makeatother

\begin{document}

\title[]{Superinductor-based ultrastrong coupling in a superconducting circuit} %Title of paper

\author{Alba Torras-Coloma}
\affiliation{Institut de Física d’Altes Energies (IFAE), The Barcelona Institute of Science and Technology (BIST), Bellaterra (Barcelona) 08193, Spain}
\affiliation{Departament de Física, Universitat Autònoma de Barcelona, 08193 Bellaterra, Spain}

\author{Luca Cozzolino}
\affiliation{Institut de Física d’Altes Energies (IFAE), The Barcelona Institute of Science and Technology (BIST), Bellaterra (Barcelona) 08193, Spain}
\affiliation{Departament de Física, Universitat Autònoma de Barcelona, 08193 Bellaterra, Spain}

\author{Ariadna Gómez-del-Pulgar-Martínez}
\affiliation{Institut de Física d’Altes Energies (IFAE), The Barcelona Institute of Science and Technology (BIST), Bellaterra (Barcelona) 08193, Spain}
\affiliation{Departament de Física Quàntica i Astrofísica, Universitat de Barcelona, Barcelona 08028, Spain}

\author{Elia Bertoldo}
\affiliation{Institut de Física d’Altes Energies (IFAE), The Barcelona Institute of Science and Technology (BIST), Bellaterra (Barcelona) 08193, Spain}

\author{P. Forn-Díaz}
\affiliation{Institut de Física d’Altes Energies (IFAE), The Barcelona Institute of Science and Technology (BIST), Bellaterra (Barcelona) 08193, Spain}
\affiliation{Qilimanjaro Quantum Tech SL, Barcelona, Spain}
\email{pforndiaz@ifae.es}

\date{\today}

\begin{abstract}
We present an ultrastrong superinductor-based coupling consisting of a flux qubit galvanically coupled to a resonator. The coupling inductor is fabricated in granular Aluminum, a superinductor material able to provide large surface inductances. Spectroscopy measurements on the qubit-resonator system reveal a Bloch-Siegert shift of \SI{23}{\mega\hertz} and a coupling fraction of $g/\omega_r \simeq 0.13$, entering the perturbative ultrastrong coupling (USC) regime. We estimate the inductance of the coupler independently by low-temperature resistance measurements providing $L_c = (0.74\pm0.14)\,\mathrm{nH}$, which is compatible with $g/\omega \gtrsim 0.1$. Our results show that superinductors are a promising tool to study USC physics in high-coherence circuits using flux qubits with small loop areas and low persistent currents.

\end{abstract}

\pacs{}% insert suggested PACS numbers in braces on next line

\maketitle %\maketitle must follow title, authors, abstract and \pacs

Understanding the interaction between light and matter in superconducting quantum circuits~\cite{blais2004cavity, wallraff2004strong} is key to develop new designs for quantum computing \cite{stassi2020scalable} and quantum sensing applications~\cite{degen2017quantum}. The advantage of superconducting circuits is the tunability of device parameters to achieve light-matter coupling coefficients $g$ in a wide range of values. In particular, $g$ achieved a sizable fraction of the excitation frequency in a qubit-resonator circuit in various experiments~\cite{yoshihara2017superconducting, forn2010observation, niemczyk2010circuit}, entering the so-called ultrastrong coupling (USC) regime~\cite{ciuti2005quantum, bourassa2009ultrastrong}. This boundary is defined by interaction strengths in the range $g/\omega_r\gtrsim0.1$, where $\omega_r$ is the bare frequency of the resonator \cite{rossatto2017spectral}. Beyond this boundary, the qubit-resonator system cannot be considered as two independent bodies, giving rise to new physics with practical applications in quantum information science \cite{forn2019ultrastrong, frisk2019ultrastrong}. 

Different methods exist to achieve ultrastrong couplings in superconducting quantum circuits. Inductive ultrastrong couplings between flux qubits and resonators have been typically achieved using a shared Josephson junction \cite{niemczyk2010circuit, forn2017ultrastrong, yoshihara2017characteristic, yoshihara2017superconducting, tomonaga2021quasiparticle}. However, Josephson junctions have a number of disadvantages which can impact the coherence and properties of the system: junction losses and stray nonlinearities coming from two-level system defects \cite{bilmes2017electronic} and quasiparticles \cite{riste2013millisecond}. An alternative to Josephson junctions is to use a linear inductive coupler with sufficiently large inductance such as a superinductor. Superinductors are circuit elements exhibiting characteristic impedances of the order of the resistance quantum $R_Q\sim \SI{6.5}{\kilo\ohm}$ at the system operating frequencies while keeping a low level of losses and nonlinearities \cite{manucharyan2012superinductance}. These are very suitable properties for applications in superconducting quantum circuits. So far, superinductors have been realized by two different approaches: Josephson junction arrays \cite{masluk2012microwave, bell2012quantum, manucharyan2009fluxonium, manucharyan2012superinductance, fraudet2025direct} and disordered superconductors. Disordered superconductors, such as those based on nitrides \cite{niepce2019high, torras2024superconducting, leduc2010titanium, vissers2010low} or oxides\cite{grunhaupt2019granular, kamenov2020granular, gupta2025low}, naturally display a large sheet kinetic inductance which in some cases can go up to several $\SI{}{\nano\henry}$ per unit area. This property can be exploited to design relatively small structures exhibiting a large inductance, which is favorable for coherent dynamics given that decoherence rates scale with both the device size as well as the circulating current \cite{braumuller2020characterizing}. 

In this work, we present a superinductor-based coupling in a qubit-resonator system where the coupling is large enough to reach the USC regime. The device consists of a 3-junction flux qubit galvanically coupled to a lumped element LC resonator. The coupling inductor is made of granular Aluminum (grAl)\cite{abeles1966enhancement, grunhaupt2019granular}, a superinductor material displaying high kinetic inductance which allows the use of smaller qubit loops and lower persistent currents. The use of a superinductor material is a key design feature to reach couplings in the USC regime without compromising qubit coherence times.

\begin{figure}[!htb]
\includegraphics{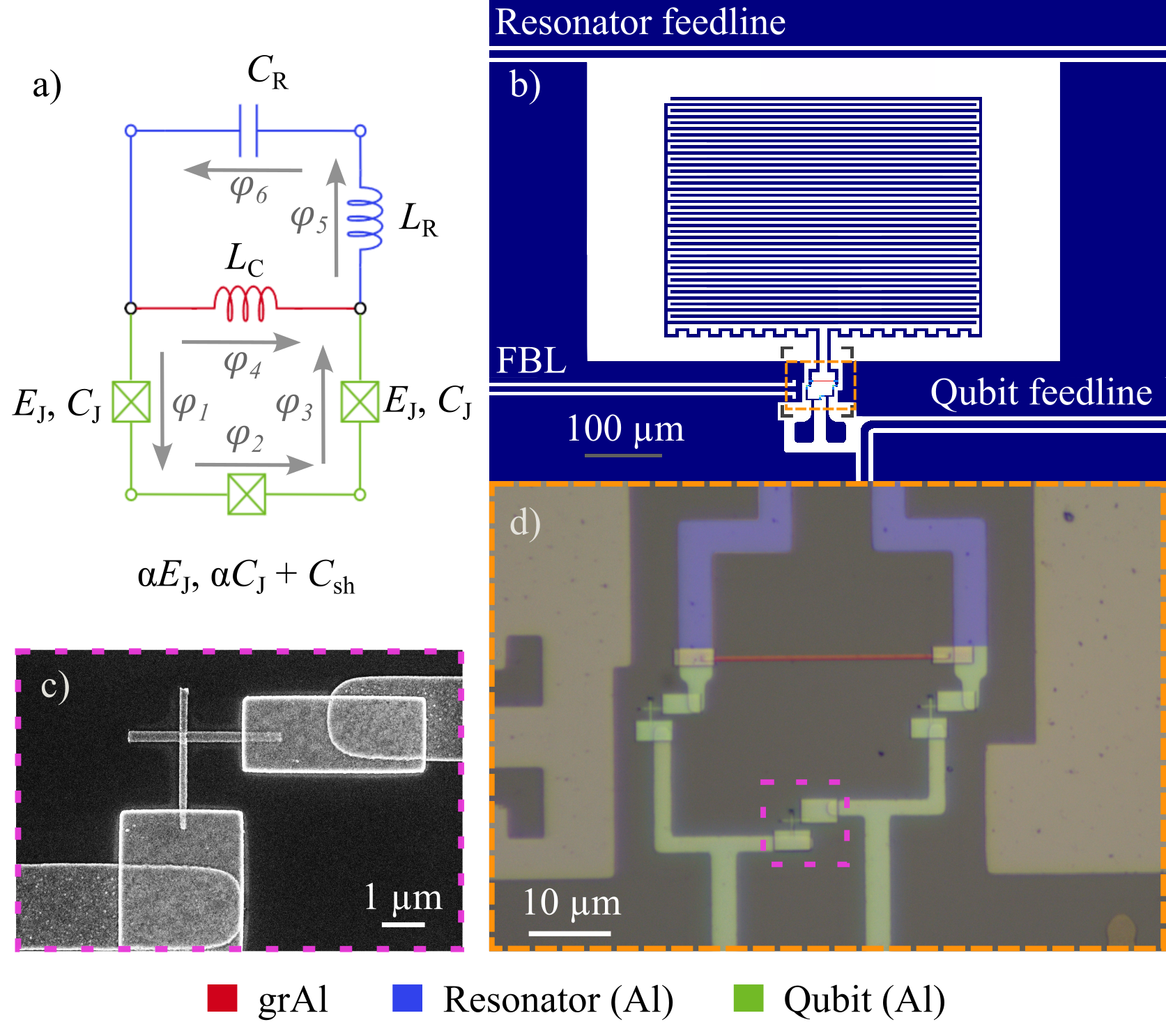}%
\caption{Qubit-resonator circuit layout.  a) Circuit schematics used to derive the full system Hamiltonian. The central junction of the qubit loop is smaller by a factor $\alpha$. The qubit and the resonator are coupled by sharing an inductor  $L_C$. b) Design layout of the system consisting of a lumped-element resonator galvanically coupled to a 3-junction flux qubit. The device contains two feedlines to selectively drive through the resonator or the qubit. c) Scanning Electron Micrograph (SEM) image of the small qubit Josephson junction. d) False-colored optical microscope image of the device. The colors indicate the materials used in each part of the chip. The grAl inductor, the qubit and the resonator are shown in red, green, and blue, respectively. The dark gray area corresponds to the Si substrate while the Al ground plane is highlighted in a lighter gray.}%
\label{fig:chip_design}
\end{figure}

The circuit schematic used in this work is presented in Fig.~\ref{fig:chip_design}a. The qubit loop is interrupted by three Josephson junctions of energy $E_J$ and capacitance $C_J$, one of which is smaller by a factor $\alpha$. By means of a shared inductance $L_c$, the qubit is galvanically coupled to an LC oscillator of inductance $L_R$ and capacitance $C_R$. As we show in Appendix~\ref{app:circuit_H}, this system can be described by the so-called Quantum Rabi Hamiltonian \cite{rabi1936process}, 

\begin{equation}
    \hat{\mathcal{H}}_{\mathrm{QRM}}  = \hbar\frac{\Omega_q}{2} \hat{\sigma}_z + \hbar \omega_r \left(\hat{a}^\dagger \hat{a}  +  \frac{1}{2}\right) + \hbar g \hat{\sigma}_x \left(\hat{a} + \hat{a}^\dagger \right),
\label{eq:qrm_hamiltonian}
\end{equation}

where $\hat{\sigma}_i$ are the Pauli operators, $g$ is the coupling strength and $\omega_r$ and $\Omega_q$ are, respectively, the resonator and qubit frequencies. Equation~(\ref{eq:qrm_hamiltonian}) can be rewritten in the flux qubit persistent current basis, 
\begin{equation}
\begin{split}
    \hat{\mathcal{H}}_{\mathrm{QRM}} & = - \frac{1}{2}\left( \epsilon \hat{\sigma}_z + \Delta \hat{\sigma}_x \right) +\\ 
    & + \hbar \omega_r \left(\hat{a}^\dagger \hat{a}  +  \frac{1}{2}\right) + \hbar g \hat{\sigma}_z \left(\hat{a} + \hat{a}^\dagger \right),
\end{split}
\label{eq:qrm_hamiltonian_qubit}
\end{equation}
where $\Delta$ is the qubit energy gap, and $\epsilon = 2I_\mathrm{p} \left(\Phi_{\mathrm{ext}} - \Phi_0/2\right)$ is the magnetic energy of the qubit with $I_\mathrm{p}$ the persistent current and $\Phi_{\mathrm{ext}}$ the external magnetic flux.

In galvanically-coupled qubit-resonator circuits such as the one in Fig.~\ref{fig:chip_design}a), an additional oscillator mode exists which is mainly determined from the combination of the coupling inductor $L_c$ and the qubit capacitances. In the perturbative USC regime, the frequency of this extra oscillator mode is at a much higher frequency than the qubit and can be adiabatically eliminated. Under this approximation, the qubit-resonator coupling strength can be estimated using~\cite{newarticle}
\begin{equation}
    g \simeq \xi_{R} \frac{L_{\mathrm{eff}} I_\mathrm{p} I_{\mathrm{rms, R}}}{\hbar} = \xi_{R}\frac{L_\mathrm{eff} I_\mathrm{p}}{\hbar Z_R} \sqrt{\frac{\hbar\omega_R}{2C_R}}
    \label{eq:coupling_coef},
\end{equation}
where $L_{\mathrm{eff}}^{-1} \equiv L_R^{-1} + L_\mathrm{c}^{-1}$ is an effective inductance, $\omega_R^2\equiv(C_RL_R)^{-1}$ is the bare resonator frequency and $\xi_{R}$ is a coefficient that approaches 1 in our circuit\footnote{The coefficient takes the form: $\xi_{R}^2 = \omega_R / \omega_A$ with $\omega_A^2 = \frac{\omega_R^2 + \omega_4^2}{2} - \sqrt{\left(\frac{\Delta^2}{2} \right)^2 + \tilde{g}^4}$, $\omega_4^2 = 1 /L_{\rm eff}C_{\rm tot}$, $\Delta^2 = \omega_R^2 - \omega_4^2$, $\tilde{g}^2 = \omega_R / \sqrt{L_R C_{\mathrm{tot}}}$, and $C_{\rm tot} = \alpha C_J+C_{\rm sh}$. The details of the derivation will be provided ref.~\cite{newarticle}.}. $I_{\mathrm{rms, R}} = (\hbar \omega_R/2L_R)^{1/2}$ and $Z_R=(L_R/C_R)^{1/2}$ are, respectively, the root-mean squared current and the impedance of the bare resonator. Note that, due to the coupling inductor $L_c$, the resonator frequency in Eq.~(\ref{eq:qrm_hamiltonian}) can be approximated as $\omega_r\simeq[(L_R+L_c)/C_R]^{-1/2}$ which differs from the bare frequency $\omega_R$ (see App.~A). 

In Fig.~\ref{fig:chip_design}b, we present the design layout of the device. Two feedlines capacitively couple to the resonator (top line) and to the shunt capacitor $C_{\mathrm{sh}}$ of the qubit (bottom line), allowing the extraction of information from the hybrid system through different observables \cite{magazzu2021transmission}. The circuit also contains a flux line coupling to the qubit loop. The lumped-element resonator is designed with a large capacitance to decrease its impedance, maximizing Eq.~(\ref{eq:coupling_coef}). The capacitance and inductance of the resonator are designed with values $C_R = \SI{0.74}{\pico\farad}$ and $L_R = \SI{0.9}{\nano\henry}$. The qubit consists of three Josephson junctions, where the central one is a factor $\alpha = 0.58$ smaller. 

The device is fabricated in five lithography steps on an intrinsic Si substrate. The ground plane, feedlines and resonator are fabricated using optical lithography and evaporated with \SI{50}{\nano\meter} of Al. The grAl, Josephson junctions and contacts are fabricated consecutively in independent electron-beam lithography steps. We adjust the resistance, and thus kinetic inductance, of grAl by the oxygen flow used in the evaporation process and the width of the coupling trace \cite{abeles1966enhancement}. In Table~\ref{tab:grAl_calibration}, we provide a list of values obtained for grAl evaporated at \SI{0.2}{nm/s} and different oxygen flows. For the present device, we use a flow of \SI{0.6}{sccm} for an expected sheet kinetic inductance of $\sim $10 pH/$\square$. The detailed process of fabrication is described in App.~\ref{app:fab}. Images of the resulting device at the end of the fabrication process are shown in Figs.~\ref{fig:chip_design}d and \ref{fig:chip_design}c. The false-colored regions indicate the different parts of the circuit. 

\begin{table}[!htb]
\caption{GrAl calibration for \SI{50}{\nano\meter} thick samples evaporated at \SI{0.2}{nm/s}. $R_s$ is the resistance measured at room-temperature as a function of the oxygen flow during the evaporation and $R_s'$ is the resistance after a \SI{13}{\minute} bake at \SI{200}{\celsius}.}
\label{tab:grAl_calibration}
\begin{tabular}{ccc} \hline\hline
$\mathrm{O_2}$ flow (sccm) & $R_s$ ($\mathrm{\Omega}/\square$) & $R_s'$ ($\mathrm{\Omega}/\square$) \\ \hline
0.0            &    $0.89 \pm 0.06$         &   $0.89 \pm 0.05$            \\
0.2            &    $2.96 \pm 0.20$          &   $2.75 \pm 0.19$            \\
0.4            &    $7.56 \pm 2.97$          &    $5.01 \pm 0.77$           \\
0.6            &    $17.31 \pm 2.48$          &  $14.57 \pm 1.49$             \\
0.8            &    $43.49 \pm 25.09$          &   $35.14 \pm 19.96$           \\ \hline\hline
\end{tabular}
\end{table}

We estimate the resistivity, $\rho$, and $L_k$ of the coupler from its dimensions and resistance at room temperature and $\SI{4}{\kelvin}$. The length of the grAl coupler is \SI{30}{\micro\meter} and its width, $(487\pm 15)\,\SI{ }{\nano\meter}$. The room temperature resistance of the wire is extracted by measuring one of the test structures, giving $R_{\mathrm{RT}} = (0.96 \pm 0.01)\,\mathrm{k\Omega}$. Using the length of the coupler and the measured thickness of the film, \SI{50}{\nano\meter}, we extract the room-temperature resistivity of the grAl line, $\rho_{RT} = (78.3 \pm 2.5 )\,\SI{}{\micro\ohm\centi\meter}$. In order to estimate the grAl inductance, we perform resistance measurements as a function of temperature on one of the grAl structures. We obtain a normal resistance of $R_{\mathrm{4K}} = (0.86 \pm 0.01)\,\SI{}{\kilo \ohm}$ and a critical temperature of $(1.60\pm0.31)\,\SI{}{\kelvin}$ (see App.~\ref{app:tc_lk_grAl}). Using the Mattis-Bardeen formula for complex conductivity in the local, dirty limit at low frequency ($hf\ll k_{B}T $) and in the low temperature limit ($T\ll T_\mathrm{c}$) \cite{rotzinger2016aluminium}, 
\begin{equation}
L_{k} = 0.18 \frac{ \hbar R_{\mathrm{4K}}}{k_B T_\mathrm{c}},
\end{equation}
we estimate the inductance of the grAl coupler to be $L_c = (0.74 \pm 0.14)\,\SI{}{\nano\henry}$. The estimated frequency and impedance of the resonator considering the renormalization effect of the coupling inductance $L_c$ is $\omega_r/2\pi \simeq 4.569$~GHz and $Z_R' \equiv[(L_R+L_c)/C_R]^{1/2} = \SI{47.1}{\ohm}$. The qubit gap $\Delta$ is designed close to $\omega_r$ by taking into account the renormalization effect of the coupling inductor $L_c$ on the qubit properties (see App.~\ref{app:simulation_spectrum} for more details). Using Eq.~(\ref{eq:coupling_coef}), we estimate the coupling to be $g/2\pi \simeq \SI{0.61}{\giga\hertz}$ for a qubit $I_p\simeq \SI{19.6}{\nano\ampere}$.

\begin{figure}[hbt]
         {\includegraphics[]{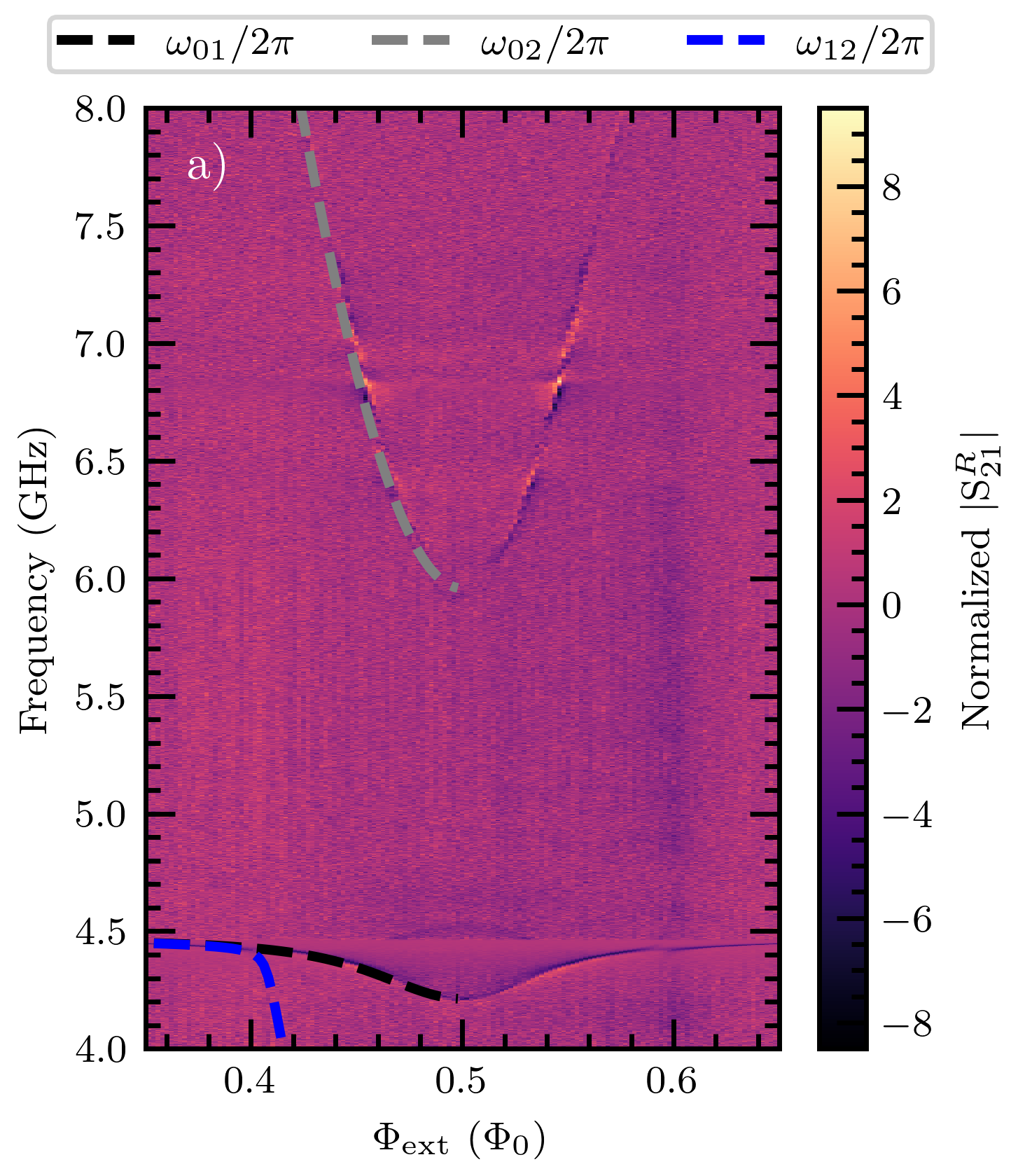}}
     \quad
         \centering
         {\includegraphics[]{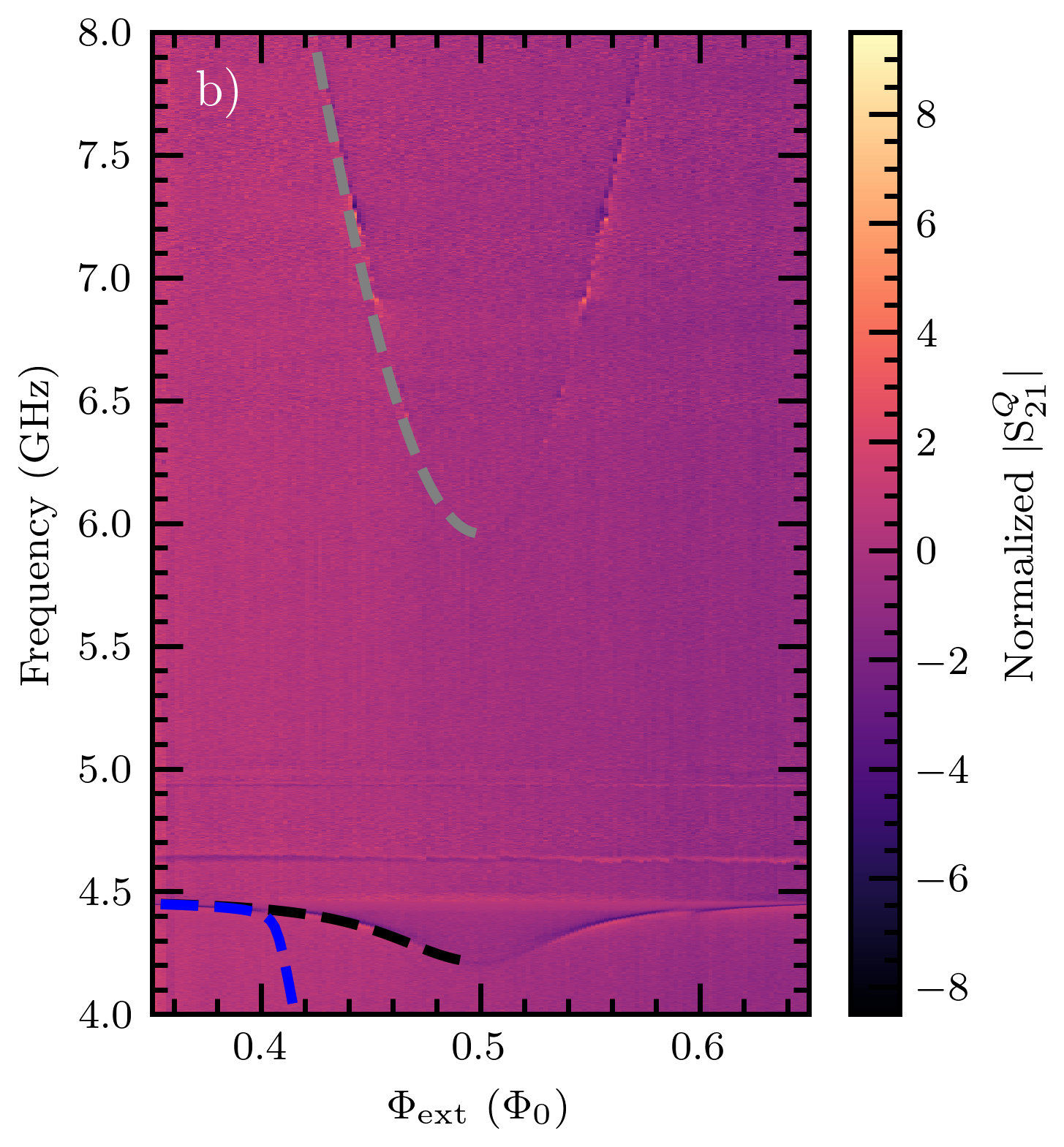}}
        \caption{Normalized transmission magnitude $|S_{21}|$ vs. flux bias $(\Phi_{\mathrm{ext}})$ obtained via single-tone spectroscopy. The expression used for the signal normalization is given in the main text. The dashed lines correspond to the fitted Quantum Rabi Model Eq.~(\ref{eq:qrm_hamiltonian}) with fitting parameters $\omega_r$, $\Delta$, $I_\mathrm{p}$ and $g$. a) Spectrum obtained measuring through the resonator feedline, $|S_{21}^R|$. b) Spectrum obtained measuring through the qubit feedline, $|S_{21}^Q|$.}
\label{fig:single_tone}
\end{figure}

The sample is packaged and mounted on the base plate of a dilution refrigerator. We measure the transmission spectrum of the system independently through the qubit ($S_{21}^Q$) or resonator ($S_{21}^R$) feedlines. An external coil is used to tune the flux through the qubit. In Fig.~\ref{fig:single_tone}, we compare single-tone transmission measurements $S_{21}^{Q, R}$ versus external flux ($\Phi_{\mathrm{ext}}$).  The magnitude of the transmission is normalized as $\left\{ |S_{21}(f_i, \Phi_{\mathrm{ext}})| - \mathrm{min}(|S_{21}|(f_i, \Phi_{\mathrm{ext}})) \right\} / \mathrm{std}(|S_{21}|(f_i, \Phi_{\mathrm{ext}}))$ where $f_i$ is the frequency trace. The normalization is used to mask the spurious box modes present in the signal (see Appendix~\ref{app:spectroscopy}). In both measurements, one can identify a high-frequency transition $\omega_{02}$ exhibiting usual flux qubit spectral features, an intermediate frequency $\omega_{01}$ mostly insensitive to flux except around the sweet spot, and a third transition $\omega_{12}$ which is much weaker and extends towards lower frequencies.

We observe minor differences between the spectra shown in Figs.~\ref{fig:single_tone}a) and b). All transitions are well defined in both plots $|S_{21}^{Q,R}|$, with a difference of $\sim\SI{5}{dB}$ outside the sweet spot when probing the system through the resonator (see the raw data provided in App.~\ref{app:spectroscopy}). In contrast, in Fig.~\ref{fig:single_tone}b) we do not observe an enhanced signal near the sweet spot for the qubit-like transition $\omega_{02}$ and a vanishing resonator-like transition $\omega_{01}$ far from the sweet spot, as theoretically expected for a qubit-driven spectrum \cite{magazzu2021transmission}. We believe that the discrepancy between theory and experiment originates from a stray coupling of the resonator to the qubit feedline. 

\begin{figure}[htb]
\centering
\includegraphics{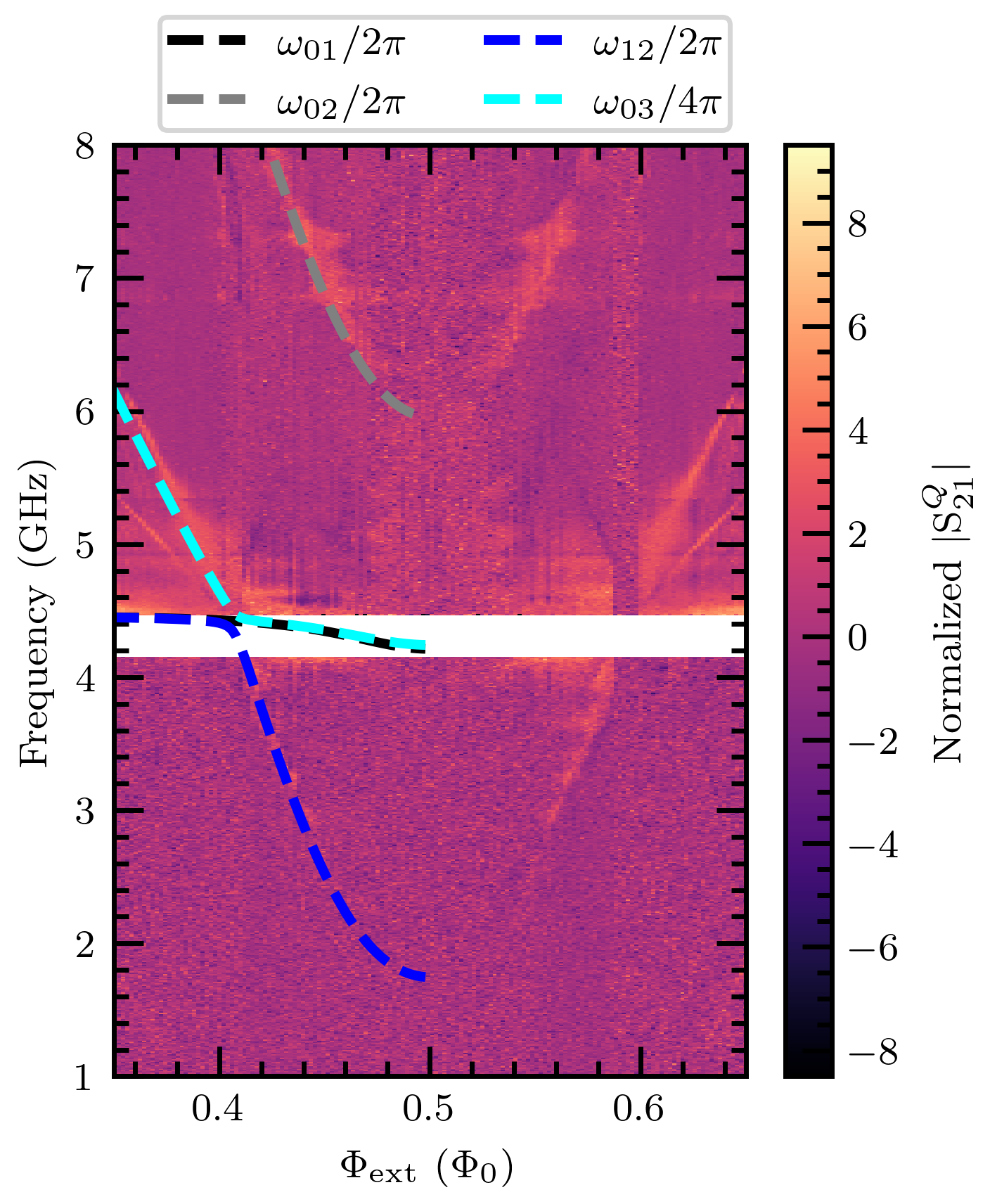}%
\caption{Normalized transmission magnitude vs. flux bias $(\Phi_{\mathrm{ext}})$ obtained via two-tone spectroscopy using the qubit feedline. The resonator tone is fixed at the frequencies defined by the $\omega_{01}$ transition measured using single-tone spectroscopy. The dashed lines correspond to the fitted Quantum Rabi Model spectrum with fit parameters $\omega_r$, $\Delta$, $I_\mathrm{p}$ and $g$. In contrast to Fig.~\ref{fig:single_tone}, we observe the two-photon transition $\omega_{03}/2$ and a second unidentified transition. }%
\label{fig:two_tone}
\end{figure}

In Fig.~\ref{fig:two_tone}, we show the normalized transmission of the two-tone spectroscopy versus flux using the values of $\omega_{01}(\Phi_{\mathrm{ext}})$ as probe. We identify four different transitions in the spectrum. $\omega_{01}$ is mostly resonator-like and is the one used for the two-tone spectroscopy measurement. $\omega_{02}$ is also present in the single-tone spectra and is mostly qubit-like. We also identify more clearly the intermediate transition $\omega_{12}$ corresponding to the exchange of an excitation between qubit and resonator. Outside the sweet spot, we also identify a two-photon qubit-like transition $\omega_{03}/2$. An additional transition is visible between \SI{4.5}{\giga\hertz} and \SI{5.5}{\giga\hertz} which seems to match to a type of three-photon blue-sideband transition $\hbar(\omega_{03}+\omega_{01})/3$.

Using the observed single-tone and two-tone transitions, the spectrum can be fitted to Eq.~(\ref{eq:qrm_hamiltonian_qubit}). The resulting parameters of the fit are $I_\mathrm{p} = (11.619 \pm 0.004)\,\SI{}{\nano\ampere}$, $\Delta / h = (5.707 \pm 0.002)\,\SI{}{\giga\hertz}$, $\omega_r/2\pi = (4.463 \pm 0.001)\, \mathrm{GHz}$, and $g/2\pi = (0.578 \pm 0.001)\,\SI{}{\giga\hertz}$. The fitted resonator frequency is compatible with the frequency $\omega_r/2\pi = \SI{4.465}{\giga\hertz}$ obtained as the saturation of $\omega_{01}(\Phi_{\mathrm{ext}})$ at high powers, similar to a punch-out measurement \cite{reed2010high}, and slightly lower than our estimate. The coupling achieved is consistent with our estimate of $L_k$ with Eq.~(\ref{eq:coupling_coef}). Despite the small value of $I_\mathrm{p}$, the fraction $g / \omega_r \simeq 0.13 > 0.1$ still falls in the perturbative USC regime. Increasing the coupler superinductance further can bring the coupling easily to  $g / \omega_r > 0.3$ while still achieving low persistent currents and relatively small qubit loops, which is advantageous for keeping long qubit coherence times.

The fitted parameters can be used to calculate the spectrum using the Jaynes-Cummings (JC) Hamiltonian \cite{cummingsicomparison}. The difference between the QRM and JC curves in the perturbative USC regime is the Bloch-Siegert shift, $\omega_{\mathrm{BS}} = g^2 / (\omega_r + \omega_q)$, an effect of the counter-rotating terms \cite{forn2010observation}. $\omega_{\mathrm{BS}}$ is maximum at the sweetspot with a value of \SI{23}{\mega\hertz} for the $\omega_{01}$ transition (see Appendix~\ref{app:QRM_vs_JC} for more details). The qubit-like transition ($\omega_{02}$) suffers the same shift but in opposite direction.  $\omega_{\mathrm{BS}}$ can be used to estimate the coupling coefficient $g/2 \pi\simeq\SI{0.48}{\giga\hertz}$ and the fraction $g/\omega_r \simeq 0.11$ which is consistent with the fitted parameters.  

In conclusion, we have presented a superinductor-based coupling consisting of a shared grAl wire between a flux qubit and a resonator. We show that although the persistent current of the qubit is low compared to previous USC experiments, the superinductor has enough contribution to bring the system in the USC regime ($g/\omega_r > 0.1$) while keeping qubit loop dimensions below $10^3\,\mathrm{\mu m^2}$. We validate the results by measuring and fitting the transmission spectra of our system to the Quantum Rabi model. Owing to the large inductance provided by the superinductor, the design constraints to reach the ultrastrong coupling can be relaxed. This work opens the door to new devices in the ultrastrong coupling regime with lower persistent current flux qubits and smaller qubit loops, thus leading to higher coherences. Our sequential fabrication procedure also allows one to introduce other materials as couplers, such as nitride-based superconductors \cite{torras2024superconducting, niepce2019high}, with potentially higher coherence times.

\begin{acknowledgments}
We would like to thank J.J. Garcia-Ripoll, L.~Magazz\`u and M.~Grifoni for the fruitful discussions. We acknowledge funding from the Ministry of Economy and Competitiveness and Agencia Estatal de Investigacion (RYC2019-028482-I, PCI2019-111838-2, PID2021-122140NB-C31), the European Commission (FET-Open AVaQus GA 899561, QuantERA), and program ‘Doctorat Industrial’ of the Agency for Management of University and Research Grants (2020 DI 41; 2024 DI 00004). IFAE is partially funded by the CERCA program of the Generalitat de Catalunya. This study was supported by MICIN with funding from European Union NextGenerationEU~(PRTR-C17.I1) and by Generalitat de Catalunya.
\end{acknowledgments}

\bibliography{bibliography}

\clearpage
\onecolumngrid

\section*{Appendix}
\subsection{Circuit Hamiltonian derivation}\label{app:circuit_H}

\begin{figure}[!htb]
    \centering
    \includegraphics[]{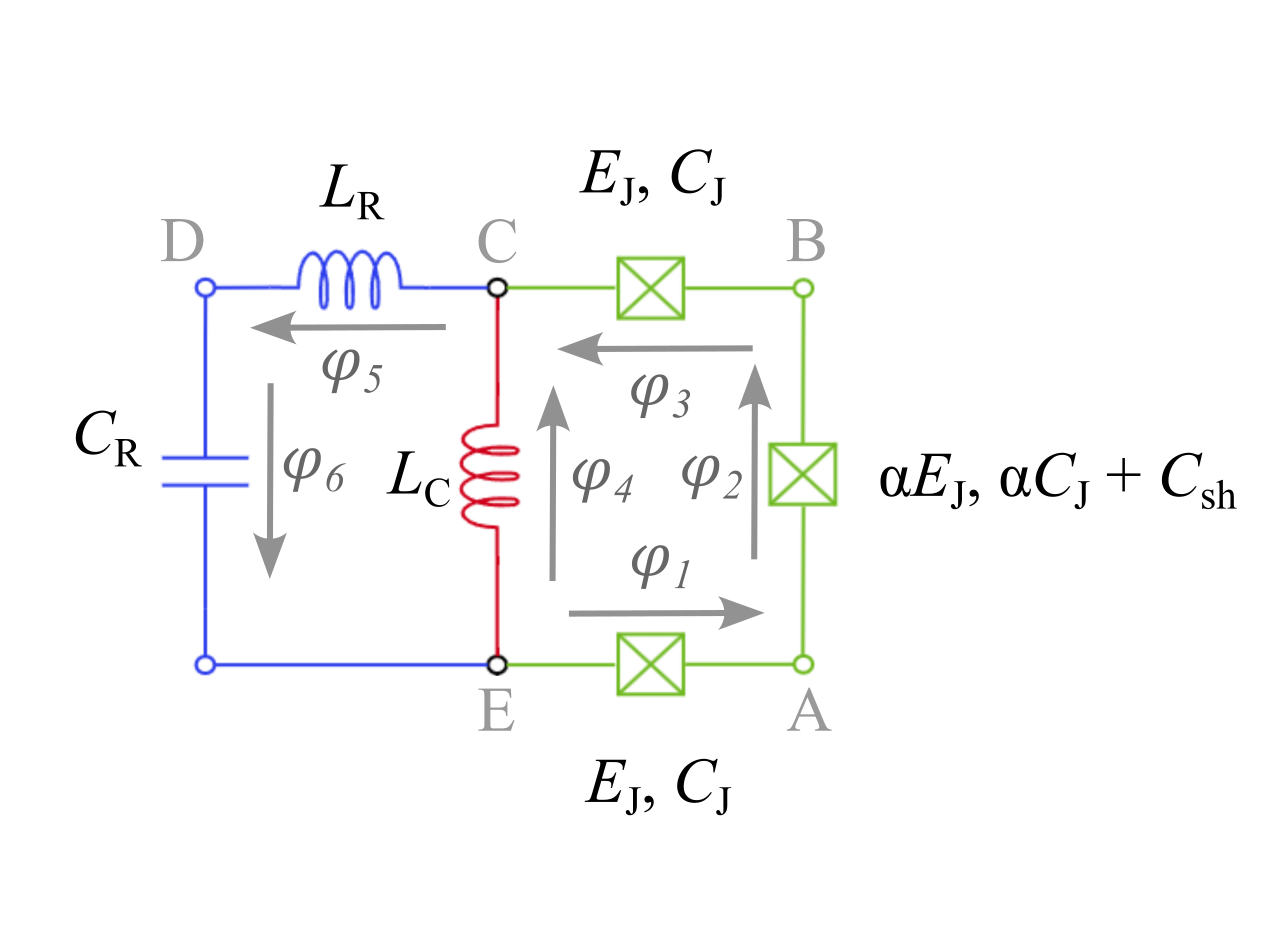}
    \caption{Circuit design consisting of a 3-junction flux qubit galvanically coupled to an LC oscillator.}
    \label{appfig:circuit}
\end{figure}

We use the circuit in Fig.~\ref{appfig:circuit} to derive the complete Hamiltonian of the system. For convenience we place the ground at node $E$. We define the branch variables of the system,

\begin{align}
    \phi_1 &= \phi_A - \phi_{\mathrm{gnd}} = \phi_A \\
    \phi_2 &= \phi_B - \phi_A \\
    \phi_3 &= \phi_C - \phi_B \\
    \phi_4 &= \phi_C - \phi_{\mathrm{gnd}} \\
    \phi_5 &= \phi_D - \phi_C \\
    \phi_6 &= \phi_{\mathrm{gnd}} - \phi_D = - \phi_D.
\end{align}

The Lagrangian in terms of the branch variables drawn in the circuit reads,

\begin{equation}
\begin{split}
    \mathcal{L} &= \frac{C_J}{2} \left(\dot{\phi}_1^2 + \Tilde{\alpha}(\dot{\phi}_4 - \dot{\phi}_1 - \dot{\phi}_3)^2 + \dot{\phi}_3^3 + \frac{C_R}{C_J}\dot{\phi}_6^2 \right) + E_J\cos{\left( \frac{2\pi}{\Phi_0}\phi_1 \right)} +  E_J\cos{\left( \frac{2\pi}{\Phi_0}\phi_3 \right)} + \\
    & + \alpha E_J\cos{\left(\left( \frac{2\pi}{\Phi_0}\right)(\phi_4 - \phi_3 - \phi_1 + \phi_{ext})\right)} - \frac{1}{2}\left(\frac{\phi_4^2}{L_c}  + \frac{(\phi_4 + \phi_6)^2}{L_R}\right),
    \label{eq:Lagrangian}
\end{split}
\end{equation}

where $L_c$ is the coupling inductance, $L_R$ and $C_R$ the resonator inductance and capacitance and $C_J$ is the Josephson junction capacitance. The central junction of the design is considered to be a fraction $\alpha$ smaller. We introduce the prefactor $\Tilde{\alpha} = \alpha + C_{sh} / C_J$ to account as well for the $C_{sh}$ capacitor of the small junction.

Using the condition for fluxoid quantization: $\phi_2 = 2\pi f - \phi_1 - \phi_3 + \phi_4$ and introducing the changes $2\pi \phi_i / \Phi_0 = \varphi_i$ and $\phi_{ext} 2\pi / \Phi_0 = 2\pi f$, the Lagrangian of the system can be rewritten as,
\begin{equation}
\begin{split}
        \mathcal{L} &= \frac{C_J}{2} \left(\frac{\Phi_0}{2\pi} \right)^2 \left(\dot{\varphi}_1^2 + \Tilde{\alpha}(\dot{\varphi}_4 - \dot{\varphi}_1 - \dot{\varphi}_3)^2 + \dot{\varphi}_3^3 + \frac{C_R}{C_J}\dot{\varphi}_6^2 \right) + E_J\cos{\varphi_1} + E_J\cos{\varphi_3} + \\
    & + \alpha E_J\cos{\left(\varphi_4 - \varphi_3 - \varphi_1 + 2\pi f\right)} - \frac{1}{2}\left( \frac{\Phi_0}{2\pi}\right)^2\left(\frac{\varphi_4^2}{L_c}  + \frac{(\varphi_4 + \varphi_6)^2}{L_R}\right).
    \label{eq:Lagrangian_varphi}
\end{split} 
\end{equation}

Calculating the conjugate variables of the system $p_i = \frac{\partial\mathcal{L}}{\partial\dot{\varphi}_i}$ and using the Legendre transformation, we obtain the Hamiltonian of the system which can be quantized promoting the variables to quantum mechanical operators,

\begin{equation}
    \begin{split}    
    \hat{\mathcal{H}} &= \left(\frac{2\pi}{\Phi_0} \right)^2 \frac{1}{2C_J}\left( (\hat{p}_1 + \hat{p}_4)^2 + (\hat{p}_3 + \hat{p}_4)^2 + \frac{\hat{p}_4^2}{\Tilde{\alpha}} + \frac{C_J}{C_R}\hat{p}_6^2\right) - E_J\cos{\hat{\varphi}_1} -  E_J\cos{\hat{\varphi}_3} - \\
    &- E_J\cos{(\hat{\varphi}_4 - \hat{\varphi}_1 - \hat{\varphi}_3 + 2\pi f)} + \frac{1}{2L_c}\left( \frac{\Phi_0}{2\pi}\right)^2\hat{\varphi}_4^2 + \frac{1}{2L_R}\left(\frac{\Phi_0}{2\pi} \right)^2(\hat{\varphi}_6 + \hat{\varphi}_4)^2.
\end{split}
\label{eq:hamiltonian_in_p}
\end{equation}

For a system in the perturbative ultrastrong coupling regime, it is reasonable to assume that $\hat{\varphi}_4$ in Eq. (\ref{eq:hamiltonian_in_p}) is in a quasi-steady state with respect to the other variables \cite{divincenzo2006decoherence,kafri2017tunable}. This choice allows us to neglect the kinetic terms related to coupling inductance and rewrite the Hamiltonian as:
\begin{equation}
\begin{split}
    \hat{\mathcal{H}}'& =\left(\frac{2\pi}{\Phi_0}\right)^2\frac{1}{2C_J}\left[\hat{p}_1^2+\hat{p}_3^2+\frac{C_J}{C_R}\hat{p}_6^2\right]-E_J \cos(\hat{\varphi}_1)
    -E_J \cos(\hat{\varphi}_3)- \alpha E_J \cos(\hat{\varphi}_4-\hat{\varphi}_1 -\hat{\varphi_3}+2\pi f) + \\ &+ \frac{1}{2L_c}\left(\frac{\Phi_0}{2\pi}\right)^2 \hat{\varphi}_4^2+\left(\frac{\Phi_0}{2\pi}\right)^2\frac{1}{2L_R}\left(\hat{\varphi}_6+\hat{\varphi}_4\right)^2 \label{eq:Ham_red_qrg}.
\end{split}
\end{equation}
If we assume that $\hat{\varphi}_4$ is small enough, corresponding to a small phase drop across the inductor, we can expand the cosine term involving $\hat{\varphi}_4$ up to second order, resulting in
\begin{equation}
\begin{split}
    \hat{\mathcal{H}}'' &=\left(\frac{2\pi}{\Phi_0}\right)^2\frac{1}{2C_J}\left[\hat{p}^2+\frac{1}{\gamma}\hat{p}_3^2+\frac{C_J}{C_R}\hat{p}_6^2\right]-E_J \cos(\hat{\varphi}_1)
    -\gamma E_J \cos(\hat{\varphi}_3)
    - \alpha E_J \cos(-\hat{\varphi}_1 -\hat{\varphi_3}+2\pi f) + \\ &- \alpha \hat{\varphi}_4 E_J \sin(-\hat{\varphi}_1 -\hat{\varphi_3}+2\pi f)
    +\frac{1}{2L_c}\left(\frac{\Phi_0}{2\pi}\right)^2 \hat{\varphi}_4^2+\frac{1}{2L_R}\left(\hat{\varphi}_6+\hat{\varphi}_4\right)^2 \; . \label{eq:Ham_red2_qrg}
\end{split}
\end{equation}
Note that this last Hamiltonian can be divided into three terms: qubit ($\hat{\varphi}_1$ and $\hat{\varphi}_3$), resonator ($\hat{\varphi}_6$) and interaction ($\hat{\varphi}_4$). Since we assumed that $\hat{\varphi}_4$ is quasi-steady in compared with the other variables of the system, it is reasonable to impose that the potential is in its minimum with respect to the coupling inductance variable \cite{kafri2017tunable} This results into the following relation,
\begin{equation}
\varphi_4^*=\left(\frac{2\pi}{\Phi_0}\right)^2\frac{L_R L_c}{L_c+ L_R}\alpha E_J \sin(-\varphi_1 - \varphi_3+2\pi f)-\frac{L_c}{L_c+ L_R}\varphi_6 \label{eq:sol_qrg}.
\end{equation}
Replacing $\hat{\varphi}_4$ by $\hat{\varphi}_4^*$ in the interaction part of the Hamiltonian derived in (\ref{eq:Ham_red2_qrg}) we obtain,

\begin{equation}\label{Ham_ind_sep_qrg}
    \hat{{\mathcal{H}}}_{int}^* = -\frac{1}{2}\left(\frac{2\pi}{\Phi_0}\right)^2\frac{L_R L_c}{L_c+ L_R} \alpha^2 E_J^2 \sin^2(-\hat{\varphi}_1 -\hat{\varphi_3}+2\pi f) -\frac{L_c}{L_c+ L_R}\frac{\hat{\varphi}_6^2}{2L_R}+\alpha\hat{\varphi_6} E_J\sin (-\hat{\varphi}_1 -\hat{\varphi_3}+2\pi f).
\end{equation}
This last expression contains essentially corrections to the qubit and resonator energy as well as the coupling term. If we rewrite the Josephson terms using the definition of the qubit current operator $\hat{I}_q \equiv \alpha I_C\sin(-\hat{\varphi}_1-\hat{\varphi}_3+2\pi f)$ we obtain the following expression,
\begin{equation}
   \hat{\mathcal{H}}_{int}=-\frac{1}{2}\frac{L_R L_c}{L_c+ L_R}\hat{I}_q^2-\frac{1}{2}\frac{L_c L_R}{L_c+ L_R}\hat{I}^2_{r}+\frac{L_c L_R}{L_c+ L_R}\hat{I}_q\hat{I}_{r},
   \label{app:interaction_term}
\end{equation}
where we have used the definition of the current operator of an LC resonator, 
\begin{equation}\label{Irms}
    \hat{I}_{r}=\frac{\partial \hat{Q}}{\partial t}=I_{rms}(\hat{a}^\dagger+\hat{a})=\frac{1}{L_R}\frac{\Phi_0}{2\pi}\hat{\varphi}_6.
\end{equation}

From Eq.~(\ref{app:interaction_term}), we can identify the coupling term and define the coupling coefficient magnitude $g$
\begin{equation}
    g=\frac{L_c L_R}{\hbar (L_c+L_R)}|\hat{I}_\mathrm{q}|I_{rms}.
\end{equation}
Note that in the limit $L_R\gg L_c$ and the two-state approximation of the qubit where $\hat{I}_q\simeq I_p\hat{\sigma}_z$, the coupling strength reduces to
\begin{equation}
    g=\frac{L_cI_\mathrm{p}I_{rms}}{\hbar} \; ,
\end{equation}
which is the common definition of the coupling strength in the literature \cite{forn2019ultrastrong,baust2016ultrastrong}. The general case of an arbitrary $L_c$ that leads to Eq.~(\ref{eq:coupling_coef}) will be the subject of future work~\cite{newarticle}.

From Eqs.~(\ref{eq:Ham_red2_qrg}) and (\ref{app:interaction_term}), we identify the terms associated to the qubit, the resonator and the coupling, respectively:
\begin{align}
    \hat{\mathcal{H}}_q &= \left( \frac{2\pi}{\Phi_0}\right)^2 \frac{1}{2C_J} \left(\hat{p}_1^2 + \hat{p}_3^2 \right) - E_J\cos{(\hat{\varphi}_1)} - E_J\cos{(\varphi_3)} - \frac{1}{2}\frac{L_RL_c}{L_c+L_R}\hat{I}^2_p,\label{app:Hqubit} \\
    \hat{\mathcal{H}}_R &= \left(\frac{2\pi}{\Phi_0} \right)^2\frac{1}{2C_R}\hat{p}_6^2 + \left( \frac{\Phi_0}{2\pi}\right)^2\frac{\hat{\varphi}_6^2}{2L_R} - \frac{1}{2}\frac{L_RL_c}{L_c+L_R}\hat{I}_{rms}^2 = \left(\frac{2\pi}{\Phi_0} \right)^2\frac{1}{2C_R}\hat{p}_6^2 + \left( \frac{\Phi_0}{2\pi}\right)^2\frac{\hat{\varphi}_6^2}{2(L_R+ L_c)}, \\
        \hat{\mathcal{H}}_c & = \frac{L_c L_R}{L_c + L_R}\hat{I}_p\hat{I}_{rms} = \hbar g \hat{\sigma}_z(\hat{a} + \hat{a}^{\dagger}).
\end{align}
Going to the two-level approximation of the qubit and using the Fock state basis for the resonator, the complete Hamiltonian can be written as,
\begin{equation}
    \hat{\mathcal{H}}_{2L} = -\frac{1}{2}(\epsilon \hat{\sigma}_z + \Delta\hat{\sigma}_x) +  \hbar\omega_r\left(\hat{a}^\dagger\hat{a} + \frac{1}{2}\right) + \hbar g\hat{\sigma}_z(\hat{a} + \hat{a}^\dagger),
\end{equation}
where $\Delta$ is the qubit energy gap, and $\epsilon = 2I_\mathrm{p} \left(\Phi_{\mathrm{ext}} - \Phi_0/2\right)$ is the magnetic energy of the qubit. Note that both $\Delta$ and $I_p$ contain the renormalization effect of the last term in Eq.~(\ref{app:Hqubit}) due to $L_c$. Moving to the diagonal form of the qubit term, we obtain the usual form of the Quantum Rabi model,
\begin{equation}
    \hat{\mathcal{H}}_{QRM} = \hbar\frac{\Omega_q}{2}\hat{\sigma}_z + \hbar\omega_r\left(\hat{a}^\dagger\hat{a} + \frac{1}{2}\right) +  \hbar g \hat{\sigma}_x(\hat{a} + \hat{a}^\dagger).
\end{equation}
Note that the resonator frequency is different from $\omega_R = 1/(L_RC_R)^{1/2}$ and takes the form $\omega_r \simeq 1/((L_R + L_c)C_R)^{1/2}$.

\subsection{Design and numerical simulation of the spectrum}\label{app:simulation_spectrum}

The resonator is designed with $C_R = \SI{742.3}{\femto\farad}$ and $L_R = \SI{898.6}{\pico\henry}$.
The qubit consisting of three Josephson junctions and the coupling inductor is designed with $\alpha = 0.58$ and energies $E_J/h = \SI{93.46}{\giga\hertz}$ and $E_C/h = \SI{4.94}{\giga\hertz}$. We consider a coupling inductor of \SI{0.5}{\nano\henry} which brings the qubit gap at $\SI{3.57}{\giga\hertz}$. With this set of parameters we estimate the coupling strength fraction to be $g/\omega_r \approx 0.39$. Where $\omega_r$ is the resonator frequency with the effect of $L_c$. 

\begin{figure}[!hbt]
         {\includegraphics[]{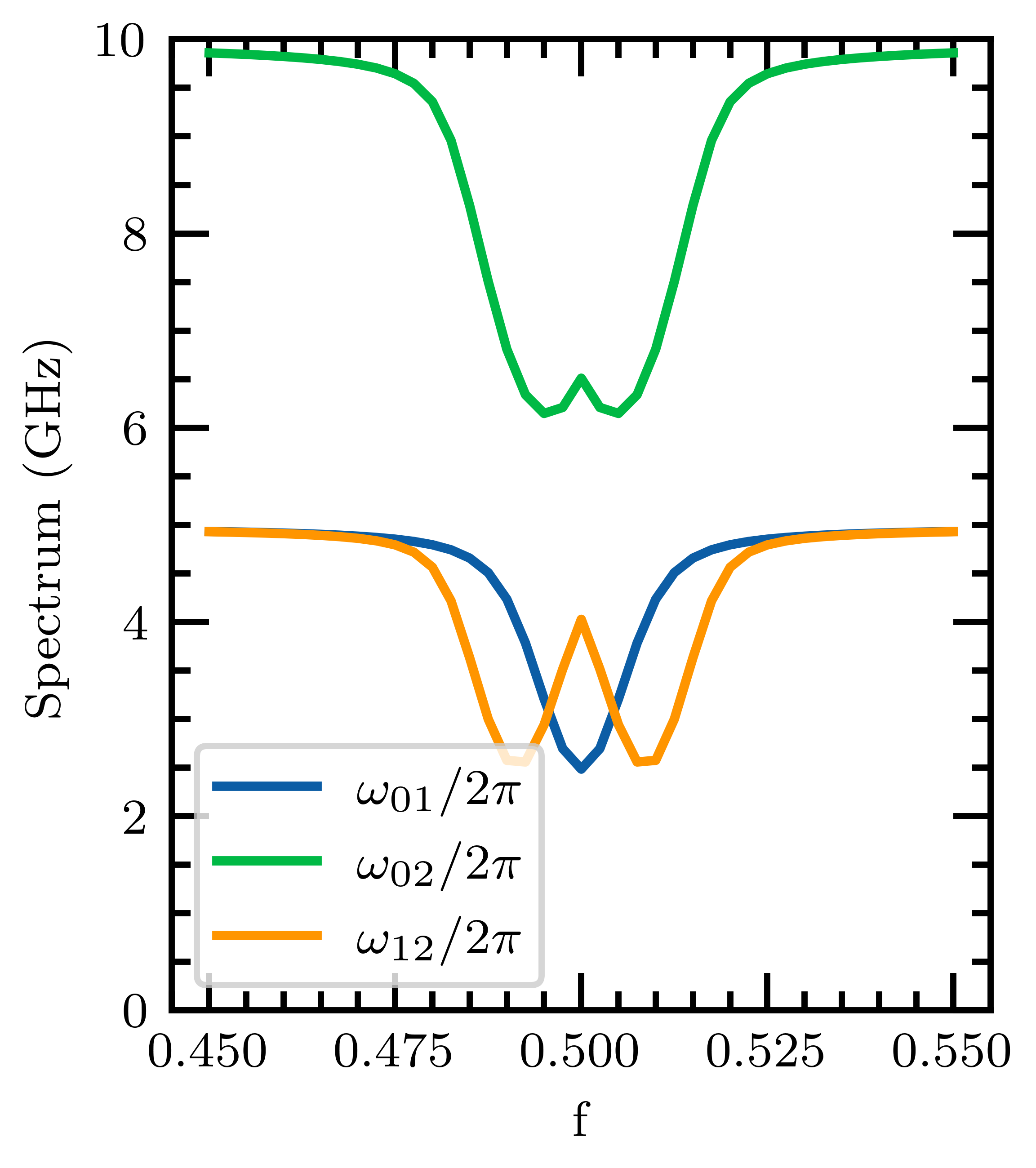}}
         %\label{appfig:design_spectrum}
     \quad
         \centering
         {\includegraphics[]{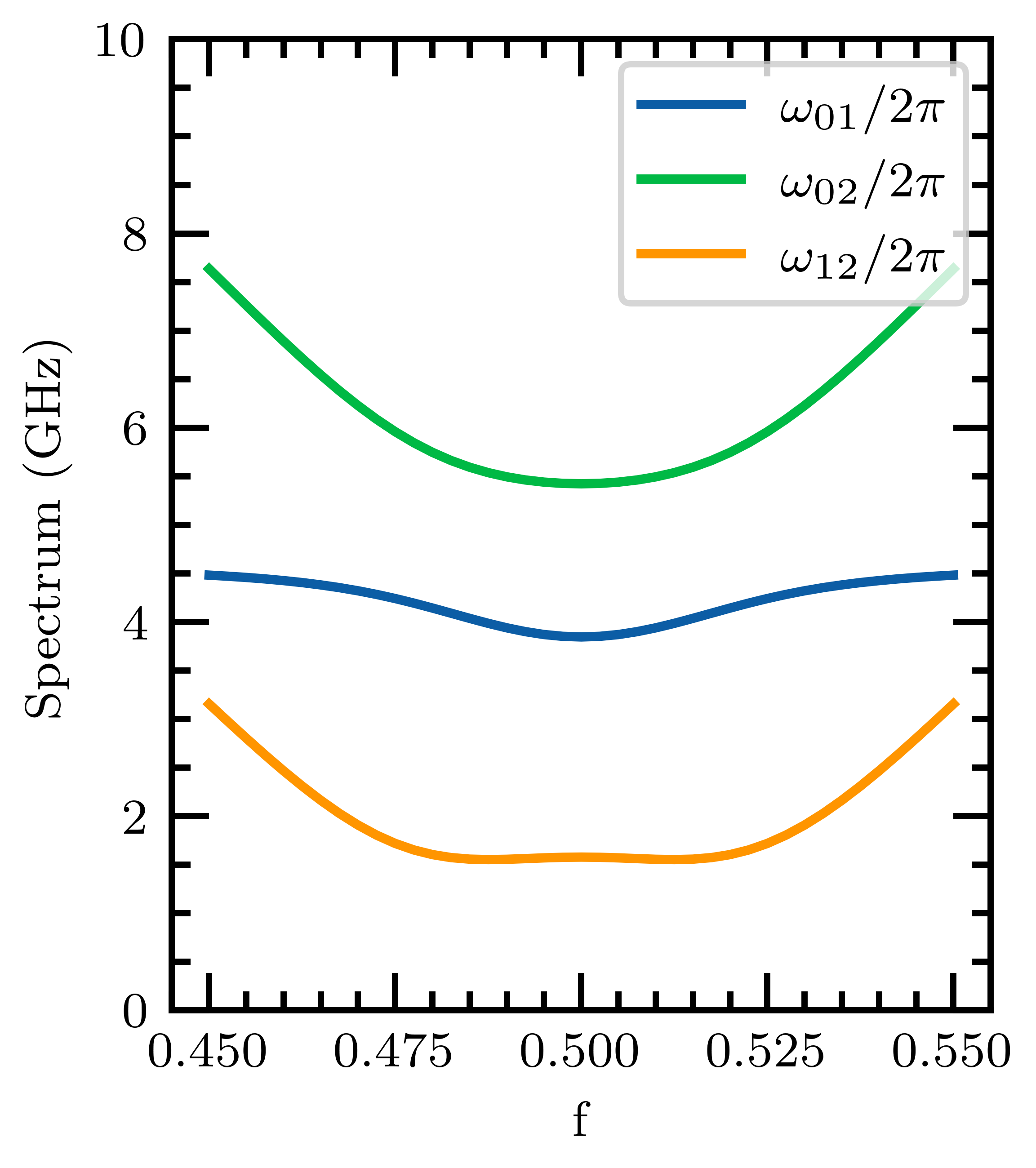}}
        \caption{Simulated spectrum of the system. Left panel: spectrum obtained with the design parameters of the circuit. Right panel: spectrum simulated with the estimated via room temperature and low-temperature measurements.}
        \label{appfig:estimated_spectrum}
\end{figure}

To diagonalize the Hamiltonian of Eq.~(\ref{eq:hamiltonian_in_p}) we choose the harmonic oscillator basis for the resonator terms and the variables associated to the branch number 4. The rest of the operators are written in the charge basis. In the left panel of Fig.~\ref{appfig:estimated_spectrum}, we show the simulated spectrum of the designed USC system and in the right panel of Fig~\ref{appfig:estimated_spectrum}, the simulated spectrum with the estimated values of $L_c=(0.74\pm 0.14)\, \mathrm{nH}$, $\alpha = (0.53\pm 0.01)$ and $J_c=(0.66 \pm 0.03)\, \mathrm{\mu A / \mu m^2}$. The resulting coupling coefficient using the estimated parameters is $g / 2\pi \approx \SI{0.61}{\giga\hertz}$ and $g/\omega_r \approx 0.15$.

\subsection{Spectroscopy data} \label{app:spectroscopy}
\begin{figure}[!hbt]
         {\includegraphics[]{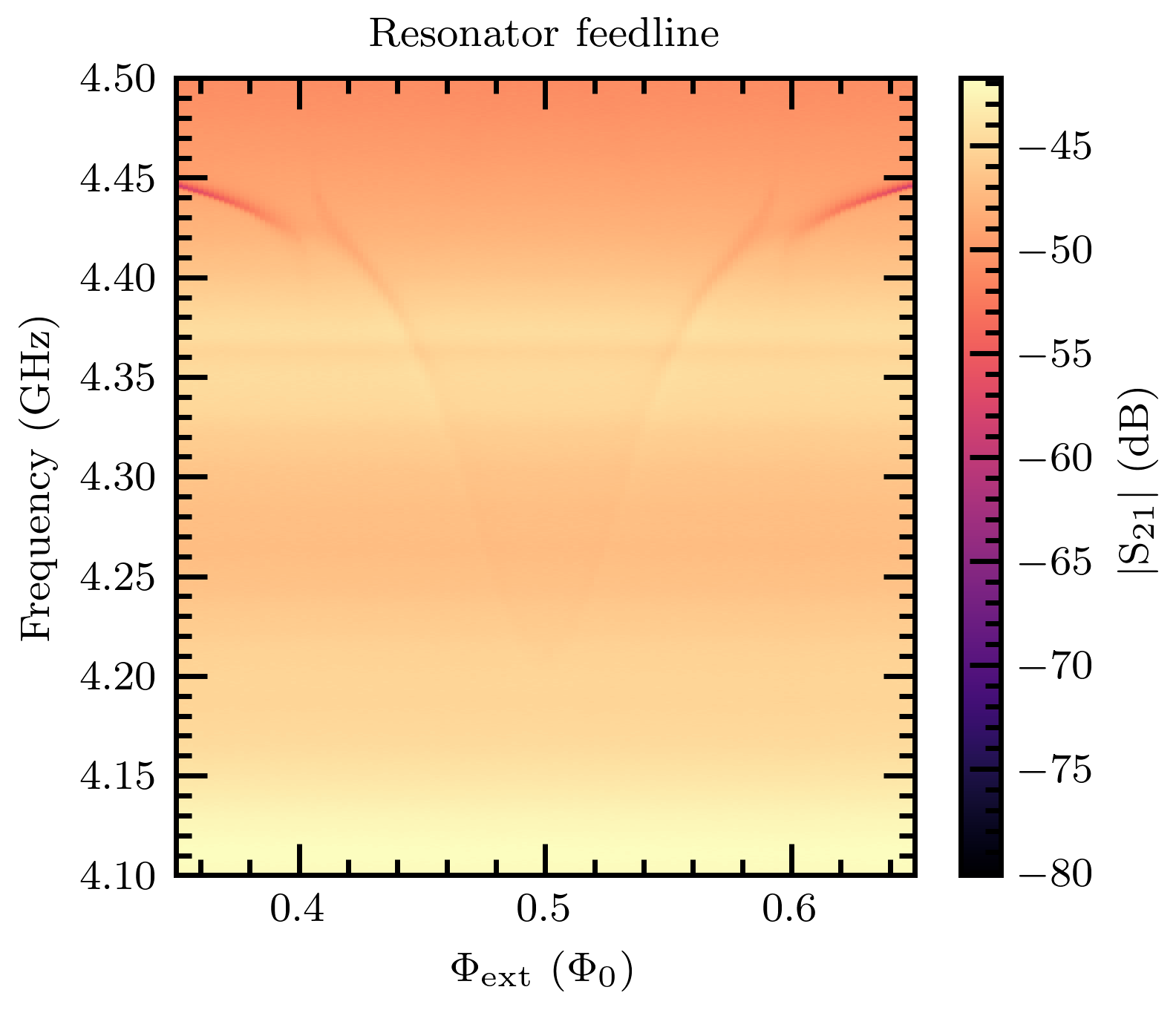}}
         %\label{appfig:design_spectrum}
     \quad
         \centering
         {\includegraphics[]{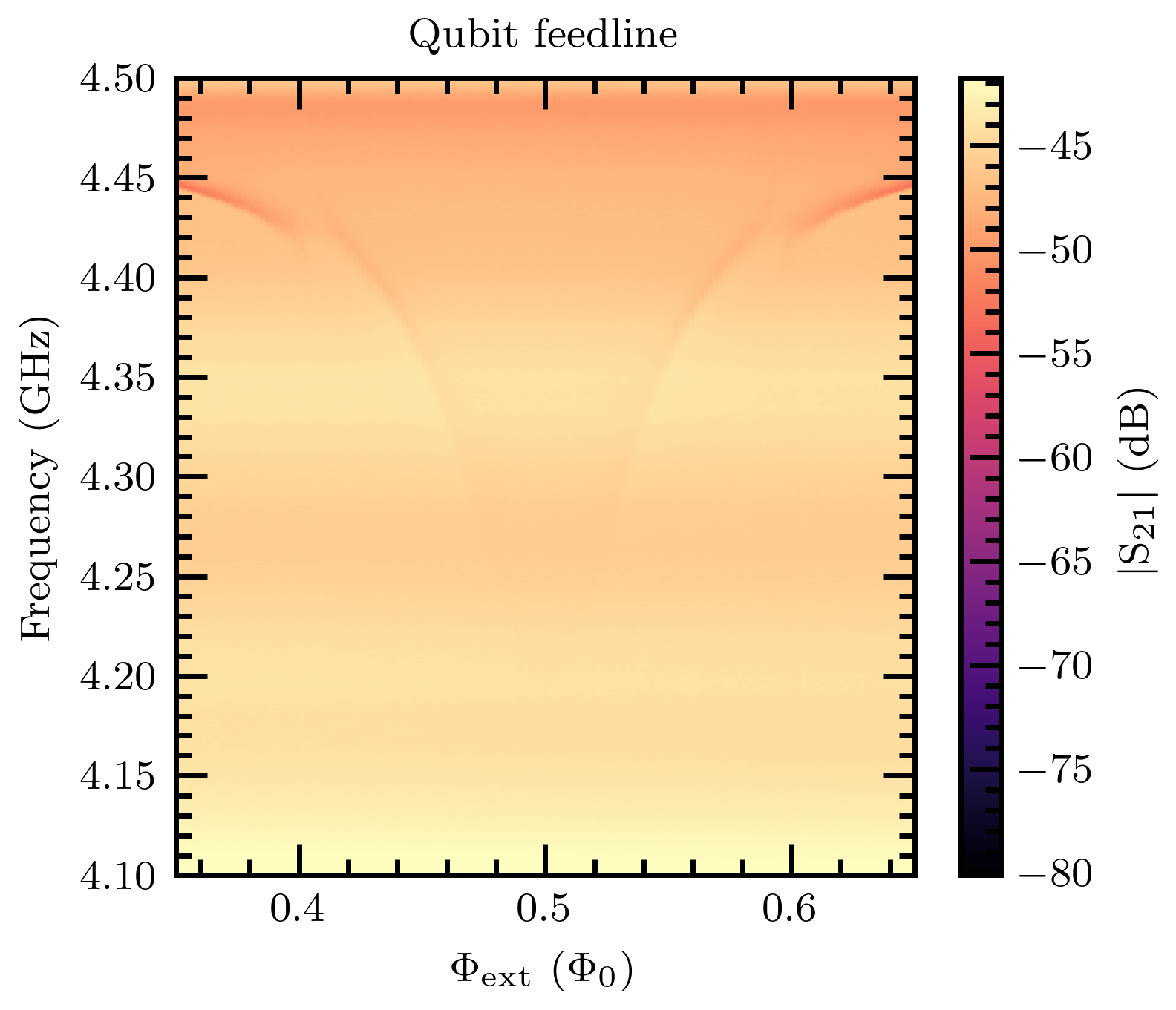}}
        \caption{Magnitude of the raw transmission obtained via single-tone spectroscopy through the resonator (left) and qubit (right) feedline.}
        \label{appfig:zoom_s21_01tr}
\end{figure}

\begin{figure}[!hbt]
         {\includegraphics[]{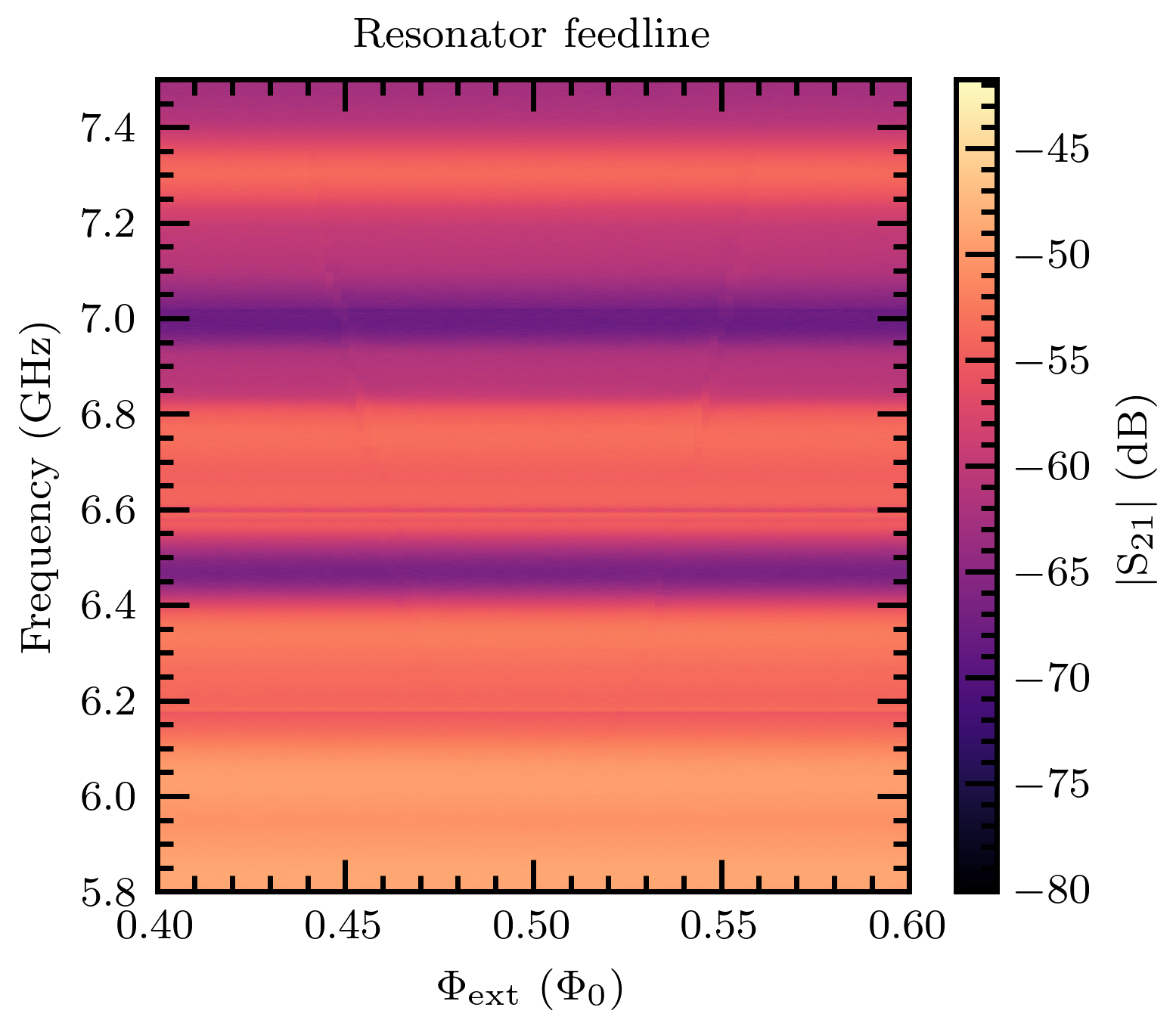}}
         %\label{appfig:design_spectrum}
     \quad
         \centering
         {\includegraphics[]{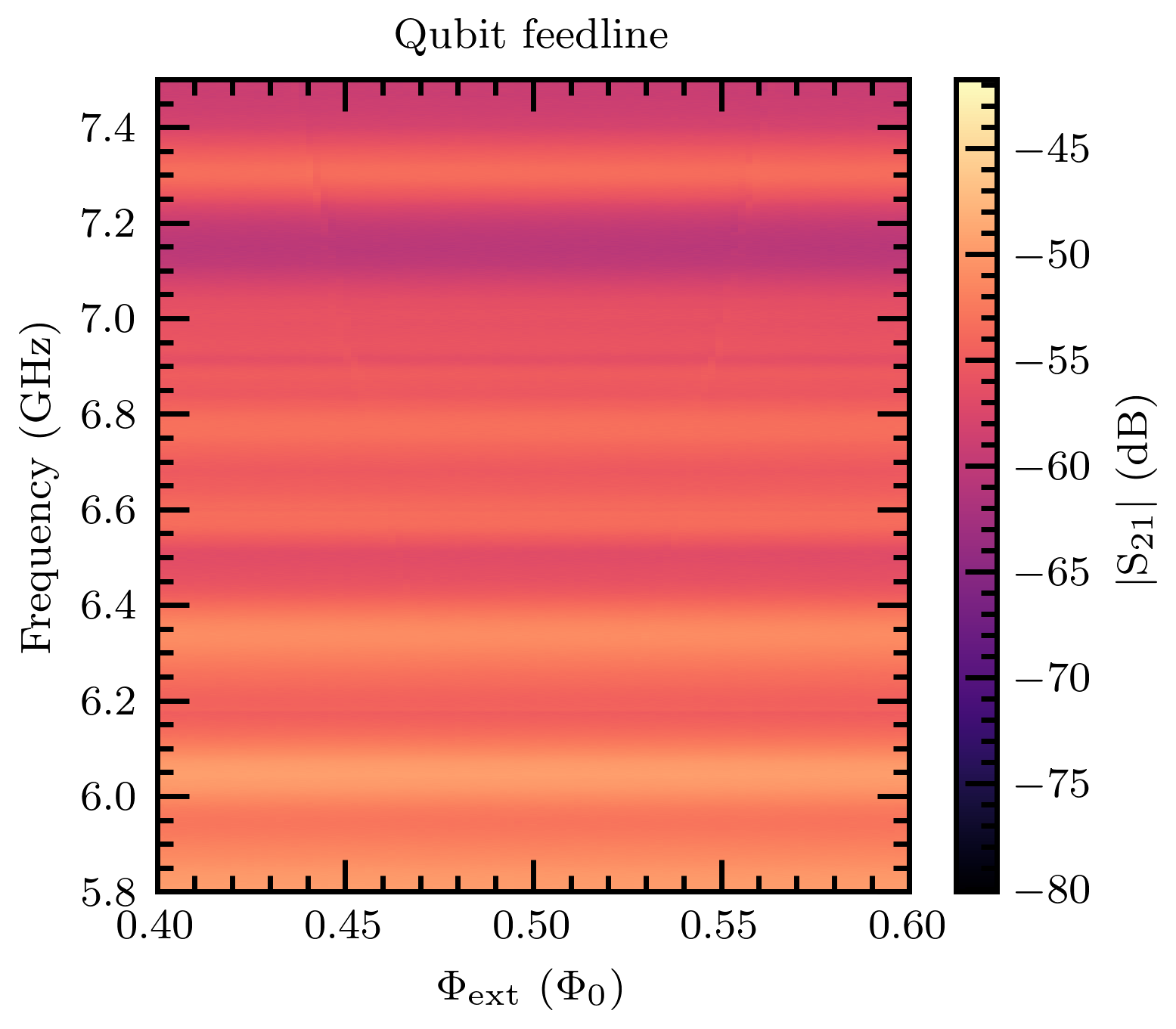}}
        \caption{Magnitude of the raw transmission obtained via single-tone spectroscopy through the resonator (left) and qubit (right) feedline.}
        \label{appfig:zoom_s21_02tr}
\end{figure}

In Fig.~\ref{appfig:zoom_s21_01tr}, we provide the raw transmission data for single-tone spectroscopy around the $\omega_{01}$ transition and Fig.~\ref{appfig:zoom_s21_02tr} for the $\omega_{02}$ transition. In the qubit feedline measurement, we included a copper piece that fills most of the empty space inside the sample box. This piece eliminated the box modes at $~ 4.36\, \mathrm{GHz}$ and $~ 6.5\, \mathrm{GHz}$ and reduced the magnitude of others spreading between $~ 6.4\, \mathrm{GHz}$ and $~ 7.5\, \mathrm{GHz}$.

\subsubsection{Quantum Rabi model and Jaynes-Cummings model} \label{app:QRM_vs_JC}
For couplings $g \ll \omega_r, \omega_q$, one can neglect the counter rotating-terms ($\hat{\sigma}_+\hat{a}^{\dagger} + \hat{\sigma}_-\hat{a}$) and use the Jaynes-Cummings (JC) Hamiltonian~\cite{cummingsicomparison}, 
\begin{equation}
    \mathcal{\hat{H}}_{JC} = \frac{\hbar\omega_q}{2}\hat{\sigma}_z + \hbar\omega_r\left(\hat{a}^\dagger \hat{a} + \frac{1}{2} \right) - \hbar g\sin{(\theta)} (\hat{\sigma}_+\hat{a} + \hat{\sigma}_- \hat{a}^\dagger) 
\end{equation}
where $\theta \equiv \arctan(\Delta / \epsilon)$, $\Delta$ is the qubit gap and $\epsilon = 2I_\mathrm{p}(\Phi_{ext} - \Phi_0 / 2)$ is the magnetic energy of the qubit. 

We calculate the JC spectrum using the fitting parameters extracted from the QRM fit: $I_\mathrm{p} = (11.619 \pm 0.004)\,\SI{}{\nano\ampere}$, $\Delta / h = (5.707 \pm 0.002)\,\SI{}{\giga\hertz}$, $\omega_r/2\pi = (4.463 \pm 0.001)\, \mathrm{GHz}$, and $g/2\pi = (0.578 \pm 0.001)\,\SI{}{\giga\hertz}$. Figure~\ref{app:QRM_vs_JC} compares the QRM results with the JC calculated spectrum. The difference for the $01$ transition at the sweetspot between both models is $\SI{23}{\mega\hertz}$ and gives an estimate of the effect of the counter-rotating terms in the form of a Bloch-Siegert shift on the qubit-resonator spectrum as discussed in the main text.
\begin{figure}[!htb]
    \centering
    \includegraphics[]{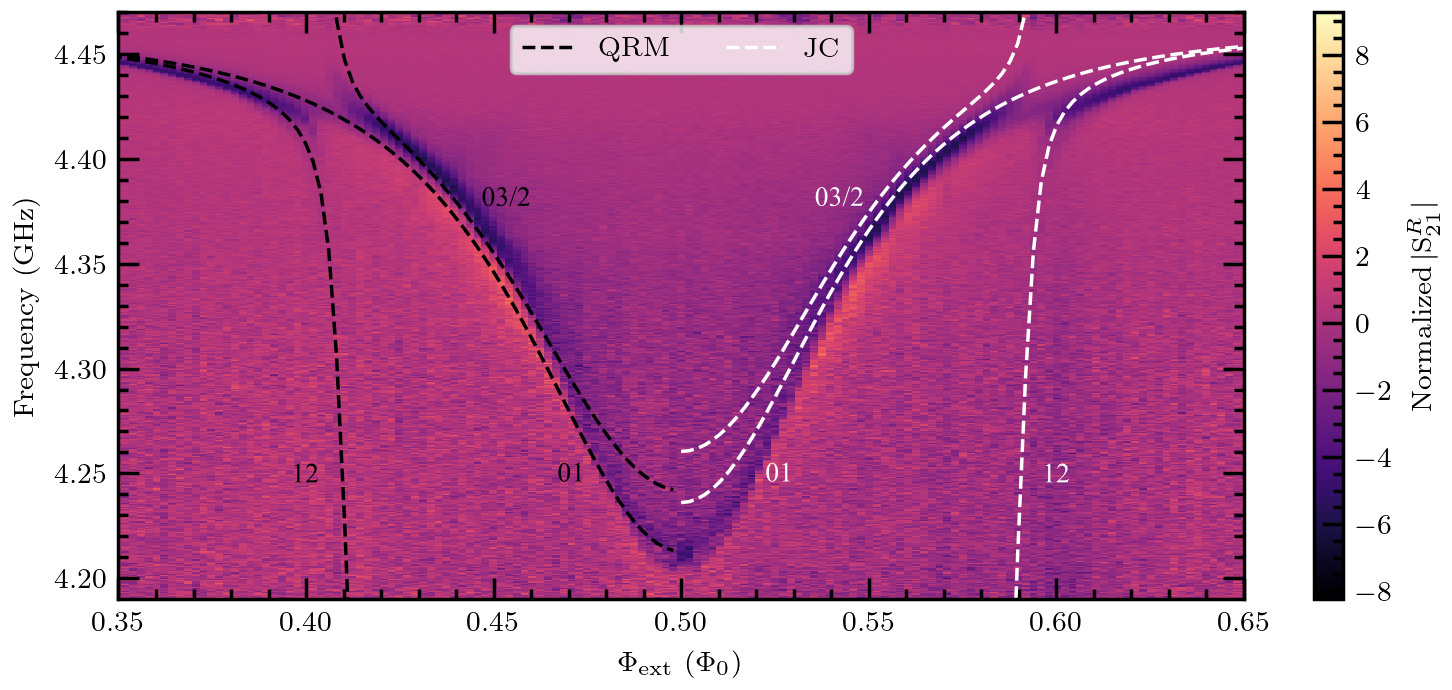}
    \caption{Zoom in around the resonator-like transition for the single-tone spectroscopy dataset presented in Fig.~\ref{fig:single_tone}. The black dashed lines are the QRM fit and the white dahsed lines are the calculated spectra using the fitting parameters and the JC Hamiltonian. The difference at the sweet spot corresponds to the Bloch-Siegert shift from the counter-rotating terms.}
    \label{appfig:QRM_vs_JC_over_single_tone}
\end{figure}

\subsection{Fabrication}\label{app:fab}

\subsubsection{Recipe for the device fabrication}

The device is fabricated on an intrinsic silicon substrate with a thickness of \SI{525}{\micro\meter}. The micron-sized structures of the design are patterned using a maskless aligner. We perform a \SI{1}{\minute} post exposure bake at \SI{110}{\celsius}. The AZ nLOF 2020 resist is later developed in AZ 726 MIF for \SI{1}{\minute} and rinsed in $\mathrm{H_2O}$.

The grAl line, Josephson junctions and contacts are patterned using a \SI{30}{keV} e-beam system. We use a double-stack resist consisting of PMGI (\SI{630}{\nano\meter} thick) and CSAR62 (\SI{240}{\nano\meter} thick) for this process. The sample is later developed in CSAR developer for \SI{60}{\second}. The reaction is stopped rinsing the sample for \SI{30}{\second} in CSAR stopper followed by a rinse in $\mathrm{H_2O}$ and IPA. The bottom layer is developed in a solution of 1:3.5 MP-351:$\mathrm{H_2O}$ for \SI{55}{\second} followed by a rinse in $\mathrm{H_2O}$ and IPA.

The depositions are performed in an Aluminum dedicated e-beam evaporator. The optical layer is evaporated vertically at \SI{0.2}{nm/s} rate. GrAl is obtained by evaporating Al at \SI{0.2}{nm/s} rate in a constant $\mathrm{O}_2$ flow of \SI{0.6}{sccm}. The junctions are evaporated at a double angle with an intermediate static oxidation of \SI{11}{\minute} at \SI{0.5}{\milli\bar}. Before depositing the contacts (\SI{100}{\nano\meter} of Al at \SI{0.2}{nm/s}) we perform an ion argon milling step to remove surface oxides.

\subsubsection{GrAl calibration}\label{app:grAl}

To evaporate grAl we pump down the load-lock chamber of our Plassys evaporator below \SI{3e-7}{\milli\bar}. Before deposition, an initial Ti evaporation is performed to decrease further the pressure. Afterwards, we change to the grAl dedicated pocket. The deposition is performed at \SI{0.2}{nm/s} under a constant flow of $\mathrm{O_2}$ gas. Changing the deposition rate and/or the flow will modify the properties of the resulting material. 

\begin{figure}[!hbt]
    \centering
    \includegraphics{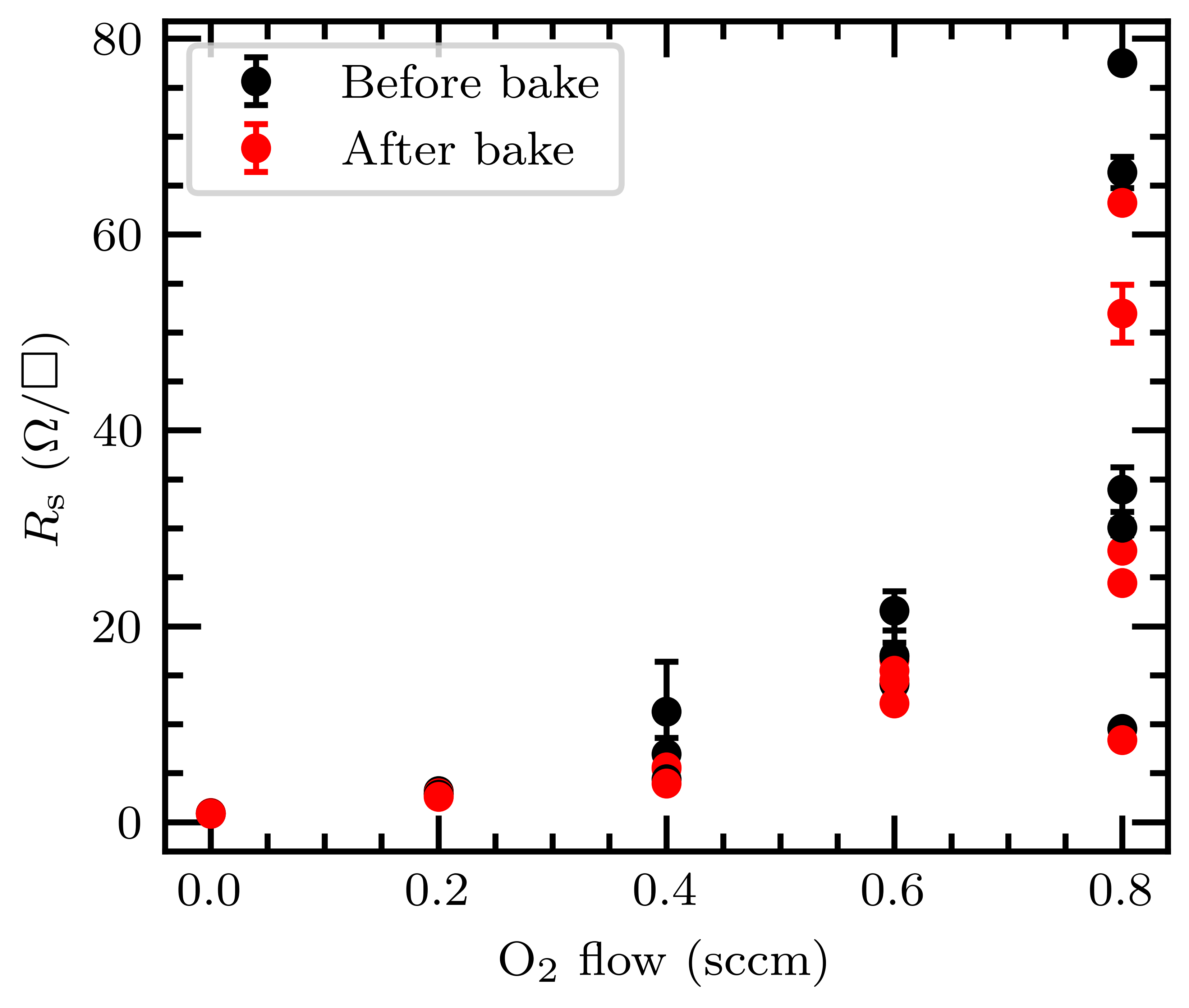}
    \caption{Sheet resistance of grAl for different oxygen flow deposition. Each sample is measured right after fabrication (before bake) and after a \SI{200}{\celsius} bake of \SI{13}{\minute}.} 
    \label{fig:gral_calibration}
\end{figure}

Figure~\ref{fig:gral_calibration} shows the sheet resistance obtained for different grAl samples with \SI{50}{\nano\meter} nominal thickness. We show the results right after fabrication (black) and after performing a \SI{200}{\celsius} bake of \SI{13}{\minute} (red). The dispersion of values for $\SI{0.8}{sccm}$ is due to problems with the rate stabilization.  

\subsection{Critical temperature of the grAl coupler}\label{app:tc_lk_grAl}

Together with the qubit device we have a set of test structures on the chip mimicking the Josephson junctions and grAl coupler. The idea of these structures is to be able to test the resistance of the different components at room temperature, but if needed, they can also be wirebonded and tested at low temperatures. 

\begin{figure}[!htb]
    \centering
    \includegraphics[]{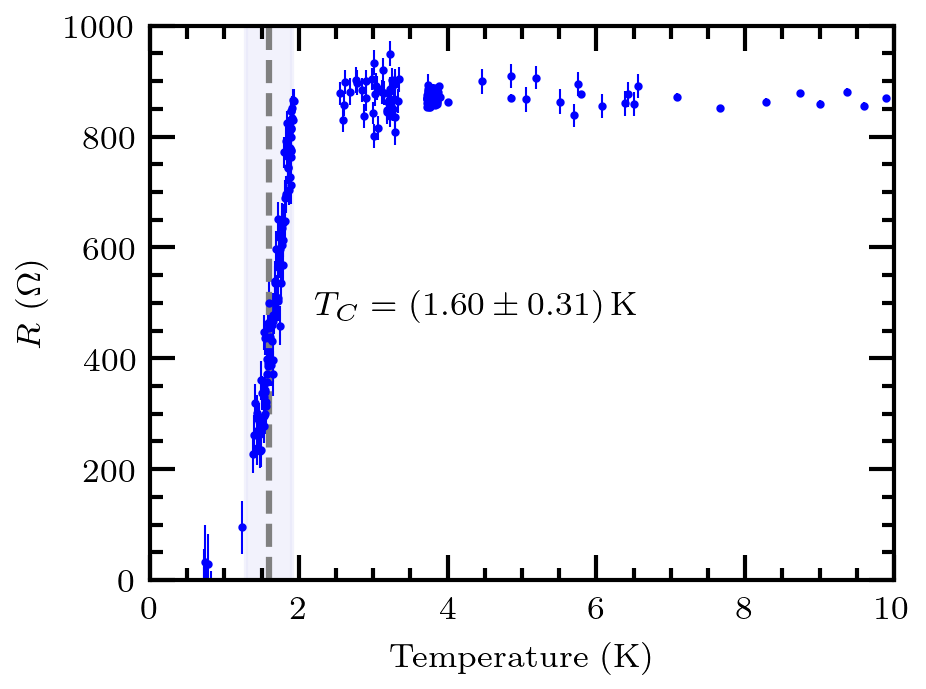}
    \caption{Resistance versus temperature curve for the grAl test structure. The dashed line indicates the $T_\mathrm{c}$ while the shadowed area gives the width of the critical temperature curve.}
    \label{appfig:grAl_Tc_plot}
\end{figure}

In order to determine the critical temperature of the grAl coupler, we wirebond in a 4-probe configuration one of grAl test structures to a commercial ceramic chip carrier. The sample is then mounted on the Still plate of a dry dilution refrigerator. We measure current-voltage (IV) curves in temperature to latter extract the resistance. Below $\SI{4}{\kelvin}$, the temperature is controlled by adding small amounts of $\mathrm{He^3/He^4}$ mixture while slowly adjusting the heaters in the refrigerator. A detailed description of the manual operation of the dilution fridge below $\SI{4}{\kelvin}$ for $T_\mathrm{c}$ measurements can be found in the Appendix of \cite{torras2024superconducting}. For low temperatures, approaching $T_\mathrm{c}$, we set low enough currents and adjust the repetition rate of the measurement to avoid excessive heat dissipation when the sample is in the normal state. Similarly, the maximum current set during the $T_\mathrm{c}$ measurement is adjusted well below the switching current of the sample. In Fig.~\ref{appfig:grAl_Tc_plot}, we provide a $R(T)$ curve for one of the test structures on the chip. We estimate the critical temperature as the point where the resistance has decreased a $50\%$ of the onset value of resistance. With this method, we obtain $T_\mathrm{c} = 1.60\pm0.31\, \mathrm{K}$, where the error in the measurement is given by half the width of the critical temperature transition defined as $\Delta T_\mathrm{c} = T_{10\%} - T_{90\%}$ where $T_{10\%}$ and $T_{90\%}$ are the temperatures where the resistance has decreased a $10\%$ and $90\%$ with respect to the onset value. 

\end{document}